\documentstyle[editedvolume,numreferences]{crckapb} 

\def\be{\begin{equation}}
\def\ee{\end{equation}}
\def\ggg{G}
\begin{opening}
\title{
SEMIHARD INTERACTIONS IN NUCLEAR COLLISIONS \protect\\
BASED ON A UNIFIED APPROACH TO HIGH ENERGY SCATTERING
}
\author{H.J. DRESCHER$^1$, M. HLADIK$^1$, S. OSTAPCHENKO$^{2,1}$, K. WERNER$^{1*}$}
\institute{ \\
$^1$ SUBATECH, Universit\'e de Nantes -- IN2P3/CNRS -- EMN, 
\hspace*{0.3cm}Nantes, France\\
$^2$ Moscow State University, Institute of Nuclear Physics, Moscow, 
\hspace*{0.3cm}Russia\\ \\
$^*$ invited speaker at the ``Workshop on Nuclear Matter in Different\\
\hspace*{0.3cm}Phases and Transitions'', Les Houches, France \\
\hspace*{0.3cm}March~31 - April~10, 1998
}
\end{opening}
\runningauthor{H.J. DRESCHER, M. HLADIK, S. OSTAPCHENKO, K. WERNER}
\runningtitle{UNIFIED APPROACH TO HIGH ENERGY SCATTERING}
\begin{document}

\section{Introduction}

Our ultimate goal is the construction of a model for interactions of two nuclei
in the energy range between several tens of GeV up to several TeV per nucleon 
in the centre-of-mass system. Such nuclear collisions are very complex, being 
composed of many components, and therefore some strategy is needed to construct
a reliable model. The central point of our approach is the hypothesis, that the
behavior of high energy interactions is universal (universality hypothesis). So,
for example, the hadronization of partons in nuclear interactions follows the 
same rules as the one in electron-positron annihilation; the radiation of 
off-shell partons in nuclear collisions is based on the same principles as the 
one in deep inelastic scattering. We construct a model for nuclear interactions
in a modular fashion. The individual modules, based on the universality 
hypothesis, are identified as building blocks for more elementary interactions 
(like $e^+e^-$, lepton-proton), and can therefore be studied in a much simpler 
context. With these building blocks under control, we can provide a quite 
reliable model for nucleus-nucleus scattering, providing in particular very 
useful tests for the complicated numerical procedures using Monte Carlo 
techniques.

\section{The Universality Hypothesis}

Generalizing proton-proton interactions,
the structure of nucleus-nucleus scattering should be as follows:
there are elementary  inelastic interactions between individual 
nucleons, realized by partonic  ``half-ladders'', where the same nucleon may 
participate in several of these elementary interactions. 
Also elastic scatterings are possible, represented by parton ladders. 
Although such 
diagrams can be calculated in the framework of perturbative QCD, there 
are quite a few problems : important cut--offs have to be chosen, one has to 
choose the appropriate evolution variables, one may question the validity of the 
``leading logarithmic approximation'', the coupling of the parton ladder to the nucleon is 
not known, the hadronization procedure is not calculable from first principles 
and so on. So there are still many unknowns, and a more detailed study is 
needed. 

Our starting point is the universality-hypothesis, saying that 
{\it the behavior of high-energy interactions is universal}.
In this case all the details of nuclear interactions can be determined by studying 
simple systems in connection with using a modular structure for modeling 
nuclear scattering. One might think of proton-proton scattering representing a 
 simple system, but this is already quite complicated considering the fact  that 
we have in general already several elementary interactions
. 
It would be desirable to study just one elementary interaction, which we refer 
to as  ``semihard Pomeron'', which will be done in the next section.

\section{The semihard Pomeron}

In order to investigate the semihard Pomeron, we turn to an even simpler 
system, namely lepton-nucleon scattering. 
A photon is exchanged between the lepton and a quark of the 
proton, where this quark represents the last one in a ``cascade'' of partons 
emitted from the nucleon. The squared diagram 
represents a parton ladder.
In the leading logarithmic approximation (LLA) the virtualities of the
partons are 
ordered such that the largest one is close to the photon \cite{rey81,alt82}. 
If we compare with proton-proton scattering, we have ordering 
from both sides with the largest virtuality in the middle, so in some sense the 
hadronic part of the lepton-proton diagram represents half of the elementary 
proton-proton diagram, and should therefore be studied first. In fact such 
statements are to some extent commonly accepted, but not carried through 
rigorously in the sense that also for example the hadronization of these two 
processes is related.

But first we investigate the so-called structure function $F_2$, related to the 
lepton-proton cross section via \cite{ell96}
\be
{d\sigma\over dx\,dQ^2}={2\pi\,\alpha^2\over Q^4x}
\left(2-2y+{y^2\over1+R}\right)\,F_2(x,Q^2)
\label{e1}
\ee
with
\be
 R = {F_L\over F_2-F_L},
\ee
and with the kinematic variables
\be
Q^2=-q^2;\quad x={q^2\over 2pq}
\ee
where $q$ and $p$ are the four-momenta of the photon the proton respectively. 
$F_L$ is the longitudinal structure function.
$F_2$ represents the hadronic part of the diagram, and is, using eq. (\ref{e1}), 
measurable. In lowest order and considering only leading logarithms of $Q^2$, 
only two diagrams contribute
, which turn out to be \cite{alt82}
\be
D_0+D_1=\sum_j e_j^2\,\delta(1-{x\over
\xi})+\sum_{ij}\int_{Q_0^2}^{Q^2}{dQ'^2\over Q'^2}
\,e_j^2{\alpha_s\over 2\pi}\,{x\over \xi}\,P_i^j({x\over \xi}),
\ee
where $P_i^j$ is the Altarelli-Parisi splitting function and $Q_0$ is
some 
cut-off of the $Q^2$ integration. The variable $\xi$ is the momentum fraction 
of the quark with respect to the proton. Assuming some proton distribution 
$f(\xi,Q^2)$ at the ``factorization scale'' $Q_0^2$, an incoherent superposition 
provides \cite{alt82}
\be
F_2(x,Q^2)=\sum_j e_j^2\,x\,f^j(x,Q^2)
\ee
with
\be
f^j(x,Q^2)=f^j(x,Q_0^2)+\sum_{ij}\int_x^1{d\xi\over\xi}
\int_{Q_0^2}^{Q^2}{dQ'^2\over Q'^2}\,f^i(\xi,Q'^2)\,{\alpha_s\over
2\pi}\,P_i^j({x\over \xi}).
\label{e2}
\ee
Iterating this equation obviously represents a parton ladder with ordered 
virtualities 
.
Strictly speaking, eq. (\ref{e2}) 
is still useless, because some of the functions $P_i^j(z)$ diverge for $z
\rightarrow 1$. However this is cured by 
considering virtual emissions, and this can be conveniently taken into account, 
by ``regularizing'' the functions $P_i^j$, which amounts to adding terms
proportional $\delta (1-z)$, 
which cures the divergence. For Monte Carlo applications it is more useful to 
proceed differently \cite{mar84}: one distinguishes between  ``resolvable'' and 
``unresolvable'' emissions. Unresolvable emissions are the virtual ones and 
emissions with very small momentum fraction ($ < \epsilon$). Then one sums over 
unresolvable emissions which provides a factor
\be
\Delta^i(Q_0^2,Q_1^2)=\exp\left\{-\sum_j\int_{Q_0^2}^{Q_1^2}{dQ^2\over
Q^2}\int_0^{1-\epsilon}
d\xi\,{\alpha_s\over 2\pi}\,P_i^j(\xi)\right\},
\ee
called  Sudakov form factor. This can also be interpreted as probability of no 
resolvable emission between $Q_0^2$ and $Q_1^2$. So the procedure amounts to only 
considering resolvable emissions, but to multiply each propagator with
$\Delta^i$. 
Based on the above discussion, we define a so-called QCD evolution function 
$E_{\rm QCD}$, representing the evolution of a parton cascade from scale $Q_0^2$ to $Q_1^2$, as
\be
E^{ij}_{\rm QCD}(Q_0^2,Q_1^2,x)=\lim_{n\to\infty}E^{(n)ij}_{\rm QCD}(Q_0^2,Q_1^2,x),
\ee
where $E_{\rm QCD}^{(n)}$ represents an ordered ladder with at most $n$ ladder rungs. This is 
calculated iteratively based on 
\begin{eqnarray}
& &E_{\rm
QCD}^{(n)ij}(Q_0^2,Q_1^2,x)=\delta(1-x)\,\delta_{ij}\,\Delta^i(Q_0^2,Q_1^2)\\
& &~~~~+\sum_k\int_{Q_0^2}^{Q_1^2}{dQ^2\over Q^2}\int_0^{1-\epsilon}{d\xi\over\xi}
\,{\alpha_s\over 2\pi}\,
E_{\rm
QCD}^{(n-1)ik}(Q_0^2,Q^2,\xi)\,\Delta^k(Q^2,Q_1^2)\,P_k^j({x\over\xi})\nonumber
\end{eqnarray}
The indices $i$, $j$, $k$ represent parton flavors.
The function $E_{\rm QCD}$ is calculated initially for discrete values of the variables, and 
later used via interpolation. In this way we are really sure to use the same QCD 
evolution for any application, it is the same in deep inelastic scattering as in 
nuclear interactions.

Next we have to determine the $x$-distribution of the first parton of the 
ladder. We expect for the momentum share of this first (the fastest) parton of 
the ladder a distribution as $1/x$, which leads to mostly small values of $x$, 
which implies on the other hand a large mass $M \sim 1/x$ \cite{lan94}. 
Such large mass objects 
are theoretically described in terms of Regge theory, the most prominent  
object at large masses being the Pomeron (I$\!$P). So, as dicussed already in
\cite{wer97}, the complete diagram 
is composed of the parton ladder, as discussed, and a soft Pomeron, given as
\be
E_{\rm soft\,I\!P}\sim x^{-\alpha_{\rm I\!P}},
\ee
and the coupling between the soft Pomeron and the nucleon, which takes the 
form
\be
C_{\rm I\!P}\sim x^{-\beta_{\rm I\!P}}.
\ee
We call $E_{\rm soft I\!P}$ also the soft evolution, to indicate that we consider this as simply 
a continuation of the QCD evolution, however, in a region where perturbative 
techniques do not apply any more. 
We consider quarks and gluons to 
be emitted from the soft Pomeron, in case of quarks we have to split the gluon 
momentum. So we have the following initial distribution,
\be
\varphi^i_{\rm I\!P}(x)=C_{\rm I\!P}\otimes E^i_{\rm soft\,I\!P},
\ee
with
\be
E^i_{\rm soft\,I\!P}=\left\{
\begin{array}{ll}
E_{\rm soft\,I\!P}&{\rm if}\,i=g\\
E_{\rm soft\,I\!P}\otimes P^q_g &{\rm if}\,i=q,\bar q
\end{array} \right. .
\ee
There is a second contribution, where a Reggeon is involved. The corresponding 
soft evolution is
\be
E_{\rm soft\,I\!R}\sim x^{-\alpha_{\rm I\!R}},
\ee
and for the Reggeon--nucleon coupling we take
\be
C_{\rm I\!R}\sim x^{-\beta_{\rm I\!R}}.
\ee
So we 
have the following initial distribution
\be
\varphi^i_{\rm I\!R}(x)=C_{\rm I\!R}\otimes E^i_{\rm soft\,I\!R},
\ee
with
\be
E^i_{\rm soft\,I\!R}=\left\{
\begin{array}{ll}
E_{\rm soft\,I\!R}&{\rm if}\,i=q,\bar q\\
0 &{\rm if}\,i=g
\end{array} \right. .
\ee
The total initial distribution is obviously
\be
\varphi^i(x)=\varphi^i_{\rm I\!P}(x)+\varphi^i_{\rm I\!R}(x),
\ee
which is by construction equal to $f^i(x, Q_0^2)$.
The distribution at scale $Q^2$ is 
\be
f^j=\sum_i\varphi^i\otimes E_{\rm QCD}^{ij}
\ee
The structure function is then calculated as :
\be
F_2(x,Q^2)=\sum_je_j^2\,x\,f^j(x,Q^2).
\ee
For $Q = Q_0$, the I$\!$P-contribution is a function which peaks at very small values 
of $x$ and then decreases monotonically towards zero for $x = 1$, the I$\!$R-contribution on the 
other hand has a maximum at large values of $x$ and goes towards zero for small 
values of $x$. The precise form of $f$ depends crucially on the 
exponent for the Pomeron-nucleon coupling, and we find a good
agreement 
for $\beta_{\rm I\!P}={1\over 2}$
.

We are now in a position to write down the expression $\ggg_{\rm semi}$ for a {\it cut
semihard Pomeron}, representing an elementary inelastic interaction in $pp$
scattering. We can divide the corresponding  diagram 
into three parts
.
We have the process involving the highest parton virtuality in the middle, and 
the upper and lower part representing each an ordered parton ladder
coupled to 
the nucleon. According to the universality hypothesis, the two latter parts are 
known from studying deep inelastic scattering, representing each the hadronic part of 
the DIS diagram
\be
f^j=\sum_i\varphi^i\otimes E^{ij}_{\rm QCD},
\ee
where $\varphi^i$  is the distribution at $Q$ = $Q_0^2$.
$E^{ij}_{\rm QCD}$ represents the evolution between $Q_0^2$ and some $Q^2$,
$f^j$
is therefore a 
function of $x$ and $Q^2$. The complete diagram is therefore, for given impact parameter $b$
and given energy squared $s$, given as
\be
\ggg_{\rm semi}=\sum_{ij}\int d\xi^+d\xi^-dQ^2\,f^i(\xi^+,Q^2)\,f^j(\xi^-,Q^2)
\,{d\sigma^{ij}_{\rm Born}\over dQ^2}(\xi^+\xi^-s,Q^2),
\ee
This may be written as
\be
\ggg_{\rm semi}=\sum_{I J}\int dx^+dx^-\,\tilde\ggg_{\rm semi}^{IJ}(x^+,x^-),
\ee
with
\begin{eqnarray}
& & \tilde\ggg_{\rm semi}^{IJ}(x^+,x^-)=C_{I}(x^+)\,C_{J}(x^-)
\,\sum_{ijkl}\,\int du^+ du^- dQ^2
\,E^k_{{\rm soft}\,I}\otimes E^{ki}_{\rm QCD}(u^+)\nonumber\\
& & ~~~~~~~~~~~~~~~~~~~~E^l_{{\rm soft}\,J}\otimes E^{lj}_{\rm QCD}(u^-)
\,{d\sigma^{ij}_{\rm Born}\over dQ^2}(u^+u^-x^+x^-s,Q^2).
\end{eqnarray}
The variables $I$ and $J$ may take the values ${\rm I\!P}$ and ${\rm I\!R}$.
Integrating ${\ggg_{\rm semi}}$ over $b$, we obtain the cross section to produce a 
pair of hard 
partons, which is measurable since the partons show up 
as hadron jets and may be reconstructed. This cross section is therefore
called jet 
cross section, given as
\be
\sigma_{\rm jet}(s)=\int d^2b\,\ggg_{\rm semi}(s,b).
\ee
So, in particular, checking  the differential cross section $d \sigma_{\rm jet}/dt$ against data provides 
an important consistency check.

In addition to the semihard Pomeron, one has to consider the expression
representing the soft Pomeron \cite{wer93}. The latter one,  $\ggg_{\rm soft}$,
is the Fourier 
transform of a Regge pole amplitude $A\sim s^{\alpha(t)}$. 
So an elementary 
inelastic intraction in an energy range of say $10$ - $10^4$ GeV is therefore written as
\be
\ggg_{\rm tot}=\ggg_{\rm semi}+\ggg_{\rm soft}.
\ee

\section{Hadron Production}
 
As discussed in the last chapter, there exist observables 
which can be calculated without detailed knowledge about hadron production. 
There exist, however, a huge variety of data concerning hadronic observables such as 
rapidity and $p_t$ spectra of different types of hadrons, multiplicity distributions 
and so on.
From a theoretical point of view, this requires some more details to be worked out. 
In addition, the numerical procedures are getting more complicated, and in 
fact, the most convenient method is provided by the Monte Carlo technique, 
which means in this context the following: for a given reaction, one writes the 
total cross section as a sum over contributions from different configurations $X$,
\be
\sigma_{\rm tot}=\sum_{X\in \cal K} \sigma(X),
\ee
such that $\sigma(X)$ is non negative and can be interpreted as probability. The sum 
has to be replaced by an integral in case of a continuous configuration space. In general, 
such a configuration $X$ represents a sum over a certain class of diagrams, because 
individual diagrams may be negative. So our work is, besides providing 
appropriate expressions for the total cross sections, also to provide algorithms 
to generate configurations $X$ according to the corresponding distributions.

The first step amounts to generating parton configurations before worrying about 
hadronisation.

Let us start again with the case of lepton-nucleon scattering
.
We first have to generate the kinematical variables $x$ and $Q^2$ according to the lepton-
proton cross section,
\be
{d\sigma_{lp}\over dx\,dQ^2}={2\pi\,\alpha^2\over Q^4x}
\left(2-2y+{y^2\over1+R}\right)\,F_2(x,Q^2).
\ee
Then a Pomeron or Reggeon type of coupling is chosen, with probabilities $F_2^{\rm I\!P}/F_2$ and 
$F_2^{\rm I\!R}/F_2$ respectively. We have
\be
F_2^{\rm I\!P/I\!R}=x\,\sum_{ij}\,e_j^2\,\varphi^{i}_{\rm I\!P/I\!R}\otimes 
E_{\rm QCD}^{ij}.
\ee
$E_{\rm QCD}$ can be written as
\be
E_{\rm QCD}^{ij}(Q_0^2,Q^2,z)=\delta(1-z)\,\delta_{ij}\,\Delta^i(Q_0^2,Q^2)
+\tilde E_{\rm QCD}^{ij}(Q_0^2,Q^2,z),
\ee
where the first term represents no parton emission and the second term at least 
one emission. The probability of no emission is therefore
\be
{1\over F_2^{\rm I\!P/I\!R}}\,x\,\sum_j e_j^2\,\varphi^j_{\rm I\!P/I\!R}(x)
\,\Delta^j(Q_0^2,Q^2),
\ee
and a configuration with no emission is generated according to this probability. 
In case of at least one emission, one generates the flavour $i$ and the fraction $x_0$ of 
the first parton of the QCD cascade according to 
\be
\sum_j {e_j^2\over x_0}\,\varphi^i_{\rm I\!P/I\!R}(x_0)\,
\tilde E_{\rm QCD}^{ij}(Q_0^2,Q^2,{x\over x_0}).
\ee
We are left with the problem of generating the cascade of partons starting 
from a parton with fraction $x_0$, flavour $i$, at scale $Q_0$, up to the 
photon vertex 
at scale $Q^2$. We have
\begin{eqnarray}
& & \tilde E_{\rm QCD}^{ij}(Q_0^2,Q^2,z)= \\
& & ~~~\sum_k\int{dQ_1^2\over Q_1^2}\int{dz_1\over z_1}\,
\Delta^i(Q_0^2,Q_1^2)\,{\alpha_s\over 2\pi}\,P_i^k(z_1)
\,E_{\rm QCD}^{kj}(Q_1^2,Q^2,{z\over z_1}),\nonumber
\end{eqnarray}
and therefore $k$, $z_1$, and $Q_1^2$ of the next emission are generated
according to 
\be
\sum_j e_j^2\,
{\Delta^i(Q_0^2,Q_1^2)\over Q_1^2}\,{1\over z_1}\,
{\alpha_s\over 2\pi}\,P_i^k(z_1)
\,E_{\rm QCD}^{kj}(Q_1^2,Q^2,{z\over z_1}),
\ee
and so on. This completed the description of the algorithm to generate parton 
configurations, based on exactly the same formulas used to calculate $F_2$ as 
discussed in the previous section.

The next step consists of generating with certain probabilities hadron 
configurations, starting from a given parton configuration. We cannot calculate 
those probabilities  within QCD, so we simply provide a recipe, the so-called 
string model. The first step consists of mapping a partonic configuration 
into a string 
configuration. For this purpose, we use the colour representation of the parton 
configuration: a quark is represented by a colour line, a gluon by a colour-anticolour 
pair
.
One then follows the colour flow starting from a quark via gluons, as 
intermediate steps, till one finds an antiquark. The corresponding sequences
\be
q - g_1 - g_2 \ldots - g_n - \bar q
\ee
are identified with kinky strings, where the gluons represent the kinks. Such a
string 
decays into hadron configurations with the corresponding probabilities given in 
the framework of the theory of classical relativistic strings.

The above discussion of how to generate parton and hadron configurations is not 
yet complete : the emitted partons are in general off--shell and can therefore 
radiate further partons. This so called timelike radiation is taken into account 
using standard techniques. The mapping of parton to hadron configurations still 
works the same way as discussed above.

Let us now discuss the generation of parton configurations for an elementary 
proton-proton interaction represented by a semihard Pomeron. Based on
\be
\ggg_{\rm semi}=\sum_{IJ}\int dx^+dx^-\,\tilde\ggg^{IJ}_{\rm semi}(x^+,x^-),
\ee
one generates the light come momentum fractions $x^+$, $x^-$ of the ``Pomeron ends'' 
according to $\tilde\ggg$. This quantity may be written as 
\begin{eqnarray}
\tilde\ggg^{IJ}_{\rm semi}(x^+,x^-)&=&C_{I}(x^+)\,C_{J}(x^-) \\
& & \sum_{ij}\int dz^+ dz^- 
\,E^i_{{\rm soft}\,I}(z^+)\,E^j_{{\rm soft}\,J}(z^-)
\,\sigma_{\rm jet}^{ij}(z^+z^-x^+x^-s),\nonumber
\end{eqnarray}
with
\begin{eqnarray}
\sigma_{\rm jet}^{ij}(\hat s)&=&\sum_{kl}\int dw^+ dw^- dQ^2 \\
& & E_{\rm QCD}^{ik}(Q_0^2,Q^2,w^+)\,
E_{\rm QCD}^{jl}(Q_0^2,Q^2,w^-)\,
{d\sigma_{\rm Born}^{kl}\over dQ^2}(w^+w^-\hat s,Q^2)\nonumber.
\end{eqnarray}
The integrand serves as probability distribution to generate $z^+$ and $z^-$.
The two quantities $z^+$ and $z^-$ are the momentum fractions of the ``ladder ends'' 
with respect to the ``Pomeron ends''. Knowing the ladder mass 
$\hat s = z^+z^-x^+x^-s$, we have to 
generate the ladder rungs. Generalizing
the definition of $\sigma_{\rm jet}$, we define
\begin{eqnarray}
& & \sigma_{\rm jet}^{ij}(Q_1^2,Q_2^2,\hat s)=\sum_{kl}\int dw^+ dw^- dQ^2\\
& & ~~~~~~E_{\rm QCD}^{ik}(Q_1^2,Q^2,w^+)\,
E_{\rm QCD}^{jl}(Q_2^2,Q^2,w^-)\,
{d\sigma_{\rm Born}^{kl}\over dQ^2}(w^+w^-\hat s,Q^2)\nonumber.
\end{eqnarray}
and
\begin{eqnarray}
\sigma_{\rm ord}^{ij}(Q_1^2,Q_2^2,\hat s)&=&\sum_{k}\int dw^- dQ^2~~~~~~~~~~~~~~\\
& &  E_{\rm QCD}^{jk}(Q_2^2,Q^2,w^-)\,\Delta^i(Q_1^2,Q^2)\,
{d\sigma_{\rm Born}^{ki}\over dQ^2}(w^-\hat s,Q^2)\nonumber.
\end{eqnarray}
representing ladders with ordering of virtualies on both sides 
($\sigma_{\rm jet}$) or on one 
side only ($\sigma_{\rm ord}$). 
We calculate and tabulate $\sigma_{\rm jet}$ and $\sigma_{\rm ord}$
initially so that we can use them via interpolation to generate partons.
The generation of partons is done in an iterative fashion based on 
the following ladder equation: 
\begin{eqnarray}
& & \sigma_{\rm jet}^{ij}(Q_1^2,Q_2^2,\hat s)=
\sigma_{\rm Born}^{ij}(Q_1^2,Q_2^2,\hat s)\\
& & ~~~~~~~~~~+ \sum_{k}\int{dQ^2\over Q^2}\int{d\xi \over \xi}\,
\Delta^i(Q_1^2,Q^2)\,{\alpha_s\over 2\pi}\,P_i^k(\xi)
\,\sigma_{\rm jet}^{kj}(Q^2,Q_2^2,\xi\hat s)\nonumber\\
& & ~~~~~~~~~~~+ \sigma_{\rm ord}^{ij}(Q_1^2,Q_2^2,\hat s)\nonumber.
\end{eqnarray}

Our treatment of generating parton configurations for an elementary pp 
interaction is absolutly compatible with deep inelastic scattering, it is based on 
the same building blocks, in particular on the evolution functions.

Our procedure can now be used to treat the case of many semihard Pomerons, 
for $p-p$ as well as nuclear scattering. A detailed discussion will be given in a 
future publication.

This work has been funded in part by the IN2P3/CNRS (PICS 580)
and the Russian Foundation of Fundamental Research (RFFI-98-02-22024). 
We thank the Institute for Nuclear Theory at the University of Washington for
its hospitality and for partial support during the completion of this work.


\end{document}